\documentclass{article}
\usepackage{amssymb}

\title{Sliding vectors, line bivectors, and torque}
\author{William G. Faris\\
Department of Mathematics\\
University of Arizona\\
Tucson, AZ 85721 USA \\ \\
email: faris@math.arizona.edu}

\newtheorem{theorem}{Theorem}
\newtheorem{proposition}{Proposition}
\newtheorem{lemma}{Lemma}

\newcommand{\interiorproduct}{\mathbin{\lrcorner}}
\newcommand{\area}{\mathrm{area}}

\newcommand{\eb}{\mathbf{e}}
\newcommand{\ub}{\mathbf{u}}
\newcommand{\vb}{\mathbf{v}}
\newcommand{\wb}{\mathbf{w}}
\newcommand{\xb}{\mathbf{x}}
\newcommand{\yb}{\mathbf{y}}
\newcommand{\zb}{\mathbf{z}}
\newcommand{\Mb}{\mathbf{M}}
\newcommand{\zerob}{\mathbf{0}}

\begin{document}

\maketitle

\begin{abstract}
This paper is a modern exposition of  old ideas.
The setting is a Euclidian space $E$ of dimension $n$ with associated vector space $V$ of dimension $n$.
A (non-zero) sliding vector is a vector in $V$ that is free to move, but only within a line $L$ of $E$. The set of sliding vectors
has dimension $2n-1$. This set is naturally embedded in a vector space of dimension ${n +1 \choose 2}$. An element of this
vector space will be called a line bivector. Other terms used in applications are screw and wrench. There is a nice description
of line bivectors in terms of Grassmann algebra in a projective representation. It is shown that this abstract description has a
concrete realization in terms of moment functions from $E$ to bivectors over $V$. The literature in physics and engineering mainly deals with
the special case $n=3$. The results of the paper apply in this case and to its most common application, where the vectors in $V$ represent force
and the bivectors over $V$ represent torque. It concludes with a discussion of  duality, such as that of force and velocity or of torque and angular velocity. 


\end{abstract}

\section{Introduction}


A sliding vector  is a vector with a line of application. The vector is pictured as an arrow that is free to slide within its line.
The space of sliding vectors is not closed under addition, but sliding vectors are included in a larger vector space.
The purpose of this note is to give a brief description of the theory of these objects and of various ways to picture them.

Let $E$ be $n$-dimensional Euclidean space, and let $V$ be the associated $n$-dimensional vector space. Consider three  kinds of vectors.
\begin{itemize}
\item A \emph{bound vector} is a pair $(P, \ub)$, where $P$ is in $E$ and $\ub$ is in $V$.
\item A non-zero \emph{sliding vector} is a pair $(L,\ub)$ consisting of a line $L$ in $E$ together with a non-zero vector $\ub$ in $V$
that leaves $L$ invariant. There is also a zero sliding vector $(E, \zerob)$.
\item A \emph{free vector} is a vector $\ub$ in $V$.
\end{itemize}

Each bound vector $P, \ub$  determines a sliding vector $P \land \ub$. If $\ub \neq \zerob$, then the line $L$ is determined by $P$ and $\ub$.
If $Q$ is another point on the same line $L$, then $P \land \ub = Q \land \ub$. In the case when $\ub = \zerob$, every point $P$ in $E$
gives the zero sliding vector $P \land \zerob$.

Each sliding vector $P \land \ub$ determines a free
vector $\ub$. The maps are summarized by
\begin{equation}
(P,\ub) \mapsto P \land \ub \mapsto \ub.
\end{equation}

In the following the space of sliding vectors will be denoted $E \parallel V$. It may be thought of as the space of all pairs $(P,\ub)$
with $P$ in $E$ and $\ub$ in $V$, subject to a certain equivalence relation. Thus $(P,\ub)$ is equivalent to $(Q,\vb)$ if $\ub = \vb$
and the vectors $Q-P$ and $\ub$ are linearly dependent. With this notation, the maps given above send $E \times V \to E \parallel V \to V$.

The dimension of the space of bound vectors is $2n$, while the dimension of the space of free vectors is $n$.
The dimension of the set of sliding vectors is $2n-1$. This may be seen by noting that for each non-zero free vector $\ub$
the space of lines $L$ that are aligned with $\ub$ has dimension $n-1$.

Consider two non-zero sliding vectors $P \land \ub$ and $Q \land \vb$ with lines $L$ and $N$. Suppose that $L$ and $N$ intersect in a point $R$.
Then $P \land \ub = R \land \ub$ and $Q \land \vb = R \land \vb$.
It is natural to define the sum
\begin{equation}
P \land \ub + Q \land \vb = R \land \ub + R \land \vb = R \land (\ub + \vb).
\end{equation}

Suppose that $P \land \ub$ and $Q \land \ub$ are non-zero sliding vectors with lines $L$ and $N$ that are parallel.
Furthermore, suppose that $\ub + \vb \neq \zerob$. Then
$\ub = a (\ub + \vb) $ and $\vb = b (\ub + \vb)$ with $a+b = 1$. It is natural to define the sum
\begin{equation}
P \land \ub + Q \land \vb =   P \land a (\ub + \vb) + Q \land b (\ub + \vb) = (a P + b Q) \land (\ub + \vb).
\end{equation}
Here $a P + b Q$ is the weighted combination of points $P,Q$. Since $a+b = 1$ this is another point.

When $n \geq 2$ the sliding vectors do not form a vector space; the sum of sliding vectors need not be a sliding vector.
When $n=2$ there is only one way this can happen.
This is with two lines $L, N$ that are parallel and not equal
and with non-zero vectors $\ub, \vb$ with sum $\ub + \vb = \zerob$.  When $n \geq 3$ there are many pairs of lines that are not in the same plane,
so it is very common for the sum of two sliding vectors not to be another sliding vector.

Sliding vectors form part of a larger vector space. An element of this larger space will be called a
\emph{line bivector} or a \emph{screw}.  We shall see that the
term ``line bivector'' is geometrically natural. Other terms like ``screw'' and ``wrench'' may be appropriate in physical applications.
The precise definition of line bivector is given later on, but here is a brief preview.
A line bivector may be represented (not uniquely) in the form
\begin{equation}
\Mb = P \land \ub + \alpha,
\end{equation}
where $\alpha$ is a bivector built over the $n$ dimensional vector space $V$. The space of bivectors over $V$ has dimension ${n \choose 2}$.
The dimension of the space of line bivectors
is ${n+1 \choose 2}$. The collection of all sliding vectors $P \land \ub$ where $P$ and $\ub$ both vary
is not a vector space; it is a subset of the vector space of line bivectors of dimension $2n-1$.

This is  part of an old subject; the treatise that is often cited is Robert~S.~Ball, \emph{The Theory of Screws: A study in the dynamics of a rigid body},
published in 1876. However the topic is still of current interest, in particular as a tool for robotics. Much of the work is in dimension $n=3$,
where it is natural to call a line bivector a line pseudovector.
Books on statics  (Narayan and Mittal 2016) often realize line pseudovectors
as vector-pseudovector pairs that depend on a choice of reference point.
  A recent paper by Minguzzi (2013) on screw theory realizes line pseudovectors (screws) as  affine functions (moment functions).
     The affine function approach avoids the necessity of making an arbitrary choice of reference
  point. It also justifies the terminology used in the present paper; a point vector is a certain kind of affine function on $E$ whose values are vectors in $V$, while
  a line bivector is a certain kind of affine function on $E$ whose values are bivectors over $V$. Furthermore, a point vector typically
  defines a point in $E$, while a line bivector  typically defines a line in $E$ (the principal axis). The Minguzzi article gives other useful background information and is an excellent reference overall.

The present treatment is for $n$ dimensions. This was inspired by the book by Browne (2012) on Grassmann algebra.  This book
presents a projective space point of view. Our treatment of line bivectors begins with this projective space picture. Then it is shown how the affine function description is a concrete realization of the projective space picture.

 The next part of the paper deals with situations where one makes use of the scalar product on the vector space $V$. In that case one can interpret bivectors as elements of a Lie algebra, more specifically, as infinitesimal rotations. The scalar product also gives a canonical form for line bivectors, the Poinsot central axis theorem. It also gives a way to describe the situation in dimension $n=3$, where a bivector is often represented by a pseudovector. This part concludes with
 a brief sketch of the application to rigid body mechanics. The vector $\ub$ is a force vector, and the line bivector $P \land \ub + \alpha$ represents a
moment or torque about some unspecified point. Equilibrium calculations begin and end with sliding vectors, but the intermediate steps use the more general line bivectors. The paper concludes with a discussion of a relation between line bivectors and certain dual objects.

For the convenience of the reader, here is a summary list of notations. These are explained in the body of the paper. The list also includes
some terms used in physical applications.
\begin{itemize}
\item point $P$ in vector space $E$ of dimension $n$
\item vector $\ub$ in  vector space $V$ of dimension $n$ (force)
\item sliding vector $P \land \ub$ in space $E \parallel V$ of dimension $2n-1$
\item bivector $\alpha$ in vector space $\Lambda^2(V)$ of dimension ${n \choose 2}$ (couple, moment, torque)
\item point vector $t P + \ub$ in vector space $V^\bullet$ of dimension $n+1$
\item line bivector $P \land \ub + \alpha$ in vector space $\Lambda^2(V^\bullet)$ of dimension ${n+1 \choose 2}$ (screw, moment function, wrench)
\end{itemize}
There are various relations between these spaces:
\begin{itemize}
\item  $V \rightarrow V^\bullet$ is an injection.
\item  $ E \rightarrow V^\bullet$ is an injection.
\item $(P,\ub) \mapsto P \land \ub$ is a mapping $E \times V \rightarrow E \parallel V $.
\item  $E \parallel V \rightarrow \Lambda^2(V^\bullet)$ is an injection.
\item $\Lambda^2(V) \rightarrow \Lambda^2(V^\bullet)$ is an injection.
\item $P \land \ub + \alpha \mapsto \ub$ is a mapping $\Lambda^2(V^\bullet) \rightarrow V$.
\end{itemize}
Furthermore, for every point $O^*$ in $E$ there is a corresponding isomorphism $\Lambda^2(V^\bullet) \rightarrow V \oplus \Lambda^2(V)$.
given by $P \land \ub + \alpha \mapsto (\ub, (P-O^*) \land \ub + \alpha)$.

\section{Sliding vectors in the plane}

The case $n=2$ of sliding vectors  in the plane is particularly simple. In rigid body mechanics each
vector $\ub$ represents a force. Each non-zero sliding vector $P \land \ub$
represents a force applied to the point $P$.

The first condition for static equilibrium is that the sum of the force vectors is zero. This implies that the body, initially at rest, will not
begin translational notion. However there is a stronger condition: the sum of the sliding vectors is zero. This will ensure that the
body will not begin rotational motion. Here are three  examples.

\begin{itemize}
\item Triangle of forces. Three forces $\ub, \vb, \wb$ act on a rigid body at points $P, Q, R$.
The sliding vectors are $P \land \ub, Q \land \vb, R \land \wb$.
The first condition for equilibrium is that $\ub + \vb + \wb = \zerob$. The additional condition is that the three lines
all pass through the same point $S$. In this case
\begin{equation}
P \land \ub + Q \land \vb + R \land \wb = S \land \ub + S \land \vb + S \land \wb = S \land (\ub + \vb + \wb) = S \land \zerob,
\end{equation}
which is the zero sliding vector.

\item Parallel forces. The second example is relevant to the theory of the lever. Three forces $\ub, \vb, \wb$ are parallel to each other. They
act on a rigid body at points $P, Q, R$.
  Suppose that
 $\ub + \vb \neq \zerob$, and write $\ub = a (\ub + \vb)$ and $\vb = b(\ub + \vb)$, with $a+b = 1$.  Then $P \land \ub + Q \land \vb =
 (aP + bQ) \land (\ub + \vb)$.
 The condition for equilibrium is that $\ub + \vb + \wb = \zerob$ and that the fulcrum  $R = a P + b Q$ is
 on the line determined by $P,Q$ (the lever).
 Then
 \begin{equation}
 P \land \ub + Q \land \vb + R \land \wb = (a P + b Q) \land (\ub + \vb) + R \land \wb = R \land (\ub + \vb + \wb) = R \land \zerob,
 \end{equation}
giving the zero sliding vector.

\item Equal and opposite forces. The third example begins with two forces $\ub, \vb$ with $\ub +\vb = \zerob$ acting at points $P \neq Q$.
These cannot be compensated by a third force. The sum $P \land \ub + Q \land \vb$ is not a sliding vector: it is
a pseudoscalar quantity that is independent of position. This is seen  by writing it as $P \land (-\vb) + Q \land \vb = (Q-P) \land \vb$,
where $Q-P$ is the displacement vector from $P$ to $Q$. In the case $n=2$ such a product of vectors is a pseudoscalar.
\end{itemize}

When $E$ and $V$ are two-dimensional, there is an encompassing structure that is a three-dimensional vector space. It has
a one-dimensional subspace of  pseudoscalars (including zero). The sliding vectors form the rest of the vector space; they overlap with the
pseudoscalars only in the zero element.

\section{The space $\Lambda^2(V)$ of bivectors}

This section reviews the notion
of bivector, which is an algebraic construction that may be pictured in terms of equivalent parallelograms. This is
a standard notion; see for instance the chapter on multilinear algebra in MacLane and Birkhoff (1988).

Consider a vector space $V$ of dimension $n$. There is a corresponding vector space $\Lambda^2(V)$ of dimension ${n \choose 2}$. An element
of this space is called a \emph{bivector}. Given two elements $\vb$ and $\wb$ of $V$, there is an \emph{exterior product} $\vb \land \wb$
that gives a bivector. Such a bivector is called a \emph{decomposable bivector}. A general bivector is a sum of decomposable bivectors.
The bivectors have an algebra that includes an anticommutative law
\begin{equation}
\vb \land \wb = - \wb \land \vb.
\end{equation}
 In particular, $\vb \land \vb = \zerob$.
Furthermore, the exterior product is bilinear:
\begin{eqnarray}
\ub \land (s \vb + t \wb) &=& s \ub \land \vb + t \ub \land \wb \nonumber \\
(s \ub + t \vb) \land \wb &=& s \ub \land \wb + t \vb \land \wb.
\end{eqnarray}
If the vectors $\vb$ and
$\wb$ are linearly dependent, then $\vb \land \wb = \zerob$. Otherwise, the exterior product is non-zero.

It is  difficult to picture bivectors in general, but decomposable bivectors have a simple structure.
The vectors $\vb$ and $\wb$ determine a two-dimensional
subspace with an orientation. They also determine a parallelogram in this subspace. Consider another pair of vectors in this
subspace. Then their product
$(a \vb + b \wb) \land (c \vb + d \wb) = (ad-bc) \vb \land \wb$ is a scalar multiple of the original bivector. The plane is the
same, and the orientation is the same if $ad-bc>0$ and is reversed if $ad-bc<0$. So all bivectors that determine this plane
are related by a scalar multiple. This multiple can be thought of as a parallelogram area multiplication factor. In short,
a decomposable bivector may be thought of in terms of equivalent oriented parallelograms.
In general, a bivector is a sum of decomposable bivectors. In two and three dimensions every
bivector is a decomposable bivector.
The following simple lemma will be of use later on.

\begin{lemma}[bivector lemma] Suppose $\ub \neq \zerob$. If the bivector $\ub \land \wb = \zerob$, then $\wb = a \ub$ for some
scalar $a$.
\end{lemma}

The lemma is true because if $\wb$ is not a multiple of $\ub$, then $\ub$ and $\wb$ are linearly independent, and so $\ub \land \wb \neq \zerob$.

Consider a linear function $\ell$ from $V$ to scalars.
The linear function $\ell$ may be pictured by its level sets, which are
hyperplanes. Each hyperplane is
defined by a scalar value $c$; it is the  level set consisting of all $\ub$ with $\ell(\ub) = c$.
The linear function $\ell$ defines an operation called \emph{interior product}
that takes bivectors to vectors. If $\vb \land \wb$ is a decomposable bivector, then
\begin{equation}
\ell \interiorproduct (\vb \land \wb) = \ell(\vb) \wb - \ell(\wb) \vb.
\end{equation}
This has same structure as the product rule for differentiation, except that moving $\ell$ past $\vb$ introduces
a minus sign. The condition that the interior product $\ell \interiorproduct (\vb \land \wb) = \zerob$ is
that either $\ell(\vb) = \ell(\wb) = 0$, or $\vb \land \wb = \zerob$.

There is also a notion of \emph{trivector}. In particular, a
  \emph{decomposable trivector} is  a product   $\ub \land \vb \land \wb$. If the three vectors are linearly independent, then
  they define a three-dimensional subspace. All the decomposable trivectors  determining this subspace are the same up
  to a scalar multiple.
   In general, a {trivector} is
a sum of decomposable trivectors.

\begin{lemma}[trivector lemma] Suppose $\ub \neq \zerob$.
If the trivector $\ub \land \alpha = \zerob$, then $\alpha = \ub \land \wb$ for some vector $\wb$.
\end{lemma}

Take a basis $\eb_1, \ldots, \eb_n$ of $V$. Then the bivectors $\eb_i \land \eb_j$ with $i<j$ form a basis
of $\Lambda^2(V)$. Choose the basis with $\ub =  \eb_1$.  The general bivector is $\alpha = \sum_{i<j} c_{ij} \eb_i \land \eb_j$.
Then $\ub \land \alpha = \sum_{1<i<j} c_{ij} \eb_1 \land \eb_i \land \eb_j$.
The condition implies that  the $c_{ij} = 0$ except for $i=1$. So $\alpha =  e_1 \land \sum_{1<j} c_{1j} \eb_j = \ub \land \wb$.

Remark: When the space $V$ has a given scalar product (inner product, dot product) there are other ways to represent a bivector. For dimension $n=2$ a bivector
may be represented as a pseudoscalar (a scalar with sign depending on orientation). For dimension $n=3$ a bivector may be
represented as a pseudovector (a vector with sign depending on orientation). Similarly, for dimension $n=3$ a trivector
may be represented by a pseudoscalar. For $n=3$ the role of the bivector and trivector products are played by the
pseudovector product (cross product) and the pseudoscalar triple product (cross product combined with dot product).

\section{The space $V^\bullet$ of point vectors}

The geometric approach in the following discussion uses a projective  representation. The basic object is
a vector space $V^\bullet$  of dimension $n+1$, together with a non-trivial linear function $\ell$ from $V^\bullet$ to scalars.
This is the   \emph{projective model}.
The function $\ell$ is called the \emph{level} function.
Every hyperplane of fixed level is an affine subspace. (See the appendix of MacLane and Birkhoff (1988) for the notion of affine space.)
 An element $A$ in $V^\bullet$ will be called
a \emph{point vector}.

Let $V$ be the vector subspace where $\ell = 0$. Let $E$ be the affine subspace where $\ell = 1$. In this treatment an
element $\ub$ of $V$ will be called a vector, while an element of $P$ of $E$ will be called a point. Thus there is an injection $V \mapsto V^\bullet$
that is linear and an injection $E \mapsto V^\bullet$ that is affine.
Every linear combination of points in $V$ is in $V$.  Every linear combination of points in $E$ with coefficient sum one is in $E$.
Every linear combination of points in $E$
with coefficient sum zero is in $V$.

The space $V^\bullet$ may  be viewed in the \emph{weighted point} interpretation.
 Say that $A$ is a point vector. There are two
possibilities. If $A$ is not in $V$, then $\ell(A) = t \neq 0$, and $A = tQ$ for some point $Q$ in $E$. So the point vector $A$ may
be thought of as a point $Q$ of $E$ with a weight $t \neq 0$. The other possibility is that $A=\ub$ is a vector in $V$.
The sum $sP + t Q$ of two weighted points  with $s + t \neq 0$ is another weighted point $(s+t)R$, where $R = \frac{s}{s+t} P + \frac{t}{s+t} Q$.
If $s + t = 0$, then the sum is $sP + t Q = t (Q -P)$, a vector in  $V$.

The weighted point interpretation underlies ``mass point geometry,'' which is an attractive way of making constructions in elementary
geometry (Hausner 1998). A mass point is a weighted point $tQ$ with $t>0$, so the sum of two mass points is another mass point.

Remark: The space $V^\bullet$ may also  be viewed in the \emph{Galilean space-time} interpretation. A point vector $A$ is
 a displacement in space-time. If $A$  takes one event to another event, then $\ell(A)$ is the
 time difference between the two events. If $A$ is in $V$ with $\ell(A) = 0$,  then the length of $A$ is the distance between the two simultaneous events.
There is no natural notion of distance between events that are not simultaneous.

\begin{proposition}
Fix $P$ in $E$. Then every element $A$ of $V^\bullet$ has a unique representation
\begin{equation}
A = t P + \ub
\end{equation}
with $\ub$ in $V$.
\end{proposition}

Remark: In the lecture notes by Smith (2011) the construction giving the affine space (flat) $E$ as part of the vector space $V^\bullet$ is
called \emph{inflation}. In the terminology used there, an element of $V$ is a \emph{direction vector}, and every other vector is a \emph{point vector}.

\section{The space $\Lambda^2(V^\bullet)$ of line bivectors}

The next topic is line bivectors. The
framework used here arose in the work of Hermann Grassmann (1809--1877). His definition of
the exterior product of two vectors as a decomposable bivector is now standard. It is less well known that the exterior product
of a point and a vector is a sliding vector.

 A \emph{line bivector} is an element of the space  $\Lambda^2(V^\bullet)$ of bivectors over $V^\bullet$.
 Since $V^\bullet$
has dimension $n+1$, the dimension of the space of line bivectors must be ${n+1 \choose 2}$. (In the classical
literature a line bivector is often called a \emph{screw}. This is to emphasize that it has both a linear and a rotational
aspect.)

\begin{proposition}
Let $\Mb$ be in $\Lambda^2(V^\bullet)$. Fix $P$ in $E$. Then $\Mb$ has the representation
\begin{equation}
\Mb = P \land \ub + \alpha,
\end{equation}
where $\ub$ is in $V$ and $\alpha$ is in $\Lambda^2(V)$.
\end{proposition}

The proof of this is easy. Every element of $V^\bullet$ is of the form $tP + \zb$. Every decomposable element
of $\Lambda^2(V^\bullet)$ is of the form
\begin{equation}
(tP + \zb) \land (sP + \wb) = P \land (t \wb - s \zb) + \zb \land \wb = P \land \ub + \alpha.
\end{equation}
The sum of  elements of the form $P \land \ub + \alpha$ with fixed $P$ is also an element of this form.

This representation depends on the choice of $P$. If we take $Q = P + \zb$, then
the same line bivector has the representation
\begin{equation}
P \land \ub + \alpha = Q \land \ub + \alpha + \zb \land \ub.
\end{equation}
The change in point $P$ is compensated by the change in the bivector $\alpha$.

A \emph{sliding vector} is a  decomposable bivector  that can be put in the form $Q \land \ub$, that is, the exterior
product of a point (with weight one) in $E$ with a vector in $V$. If $\ub \neq \zerob$, then
the plane in $V^\bullet$ spanned by the point vectors $Q$ in $E$ and $\ub$ in $V$ intersects $E$ in a line $L$.
This shows that sliding vectors as defined here coincide with the sliding vectors in the introduction.
There is an obvious injection $E \parallel V \to \Lambda^2(V^\bullet)$ from the space of sliding vectors to the space of line bivectors.

The sum of two sliding vectors need not be a sliding vector. The simplest example is a sum
\begin{equation}
Q \land \ub + P \land (-\ub) = (Q-P) \land \ub.
\end{equation}
This is an exterior product of two vectors from $V$, that is, a decomposable bivector. In this
context  a decomposable bivector is called a \emph{couple}. It is non-zero when
$Q-P$ and $\ub$ are linearly independent.

On the other hand, the sum of two sliding vectors is always a sliding vector plus a couple. In fact,
\begin{equation}
Q \land \ub + P \land \vb = P \land (\ub + \vb) + (Q-P) \land \ub .
\end{equation}

\begin{proposition} Fix $P$ in $E$. Every point $Q$ in $E$ has the form $Q = P + \zb$, where $\zb$ is in $E$.
Every sliding vector $Q \land \ub$ has the form
\begin{equation}
Q \land \ub = P \land \ub + \zb \land \ub.
\end{equation}
A sliding vector with line through $Q$ is the sum of a sliding vector with line through the given $P$ with a decomposable bivector having the vector
part as a factor.
\end{proposition}

A line bivector $\Mb = P \land \ub + \alpha$ has two important invariants.
\begin{itemize}
\item The \emph{vector invariant} is $ \ub$.
\item The \emph{trivector invariant} is $\ub \land \alpha$.
\end{itemize}

The vector invariant $\ub$ depends only on the line bivector $\Mb$. This is because $\ub = \ell \interiorproduct \Mb$.
This may be seen by computing
 $\ub = \ell \interiorproduct \Mb = \ell \interiorproduct ( P \land \ub) + \ell \interiorproduct \alpha$. Since $P$ is at level 1 and $\ub$ is at level 0, the first term is $\ell(P)  \ub -  \ell(\ub) P = 1 \ub - 0 P = \ub$. The second term is zero, since it is created from vectors in $V$ which are all at level 0.

The trivector invariant also depends only on the line bivector $\Mb$. This is because $\ub \land \alpha = \ub \land \Mb$.

\begin{proposition} If  $\Mb = P \land \ub + \alpha$ is a line bivector, and if the trivector invariant $\ub \land \alpha = \zerob$, then
either $\Mb = P \land \alpha$ is a sliding vector, or $\Mb = \alpha$ is a bivector.
\end{proposition}

The proof is not difficult. If $\ub = \zerob$, then $\Mb = \alpha$. Otherwise the trivector lemma  says that
$\ub \land \alpha = \zerob$ implies $\alpha = \zb \land \ub$ for some vector $\zb$. In this case
$P \land \ub + \alpha = P \land \ub + \zb \land \ub = (P + \zb) \land \ub$ is a sliding vector.

A line bivector need not be a sliding vector. However every line bivector is a sum of at most $n$ sliding vectors. The following theorem
shows that these vectors may be taken at predetermined points.

\begin{theorem}
Let $P_0, P_1, \ldots, P_{n}$ be $n+1$ points that determine the $n$-dimensional affine space $E$. Then every line bivector may be expressed as a sum
of $n$ sliding vectors in the form $P_0 \land \ub_0 + P_1 \land \ub_1 + \cdots + P_{n-1} \land \ub_{n-1}$.
\end{theorem}

Here is a proof. Consider the basis $P_0, P_1, \ldots, P_{n-1}, P_n$ of the point vector space. Then the $P_i \land P_j$ with $i<j$ form a basis of the line bivector space. An arbitrary
line bivector may be expressed in the form
\begin{equation}
\sum_{i<j} c_{ij} P_i \land P_j = \sum_{i=0}^{n-1} \left( P_i \land \sum_{j=i+1}^n c_{ij} P_j \right) = \sum_{i=0}^{n-1} P_i \land \ub_i
\end{equation}
with $ \ub_i = \sum_{j=i+1}^n c_{ij} (P_j - P_i)$.

In three dimensions every bivector in $\Lambda^2(V)$ is decomposable. The following result shows that in this case only two sliding vectors are required.

\begin{theorem}  Every line bivector  of the form $P \land \ub + \vb \land \wb$ is the sum of two sliding vectors.
In fact,
\begin{equation}
P \land \ub + \vb \land \wb =  P \land (\ub - \wb) + (P+\vb) \land \wb.
\end{equation}
\end{theorem}

The last topic of this section is the \emph{bilinear trivector invariant}, defined for a pair $\Mb_1, \Mb_2$ of line bivectors.
This is given by
\begin{equation}
\ub_1 \Mb_2 + \ub_2 \Mb_1 = \ub_1 \land \alpha_2 + \ub_1 \land (P_2 - P_1) \land \ub_2 + \ub_2 \land \alpha_1 .
\end{equation}
This should be contrasted with the {bilinear scalar invariant} discussed in the final section. They are
closely related only when $n=3$.

\section{Representation of point vectors as affine functions from $E$ to $V$}

Again $E$ is $n$-dimensional Euclidean space, and $V$ is the $n$-dimensional space of free vectors.
The space $V^\bullet$ of point vectors may be modeled as the space of  affine functions from $E$ to $V$ of the form
\begin{equation}
O \mapsto   t (P-O) + \ub.
\end{equation}
This \emph{vector affine function} interpretation is the
\emph{point-slope form} of the equation for an affine function. The point is $P,\ub$ and the constant slope is  $-t$.
This is not a general affine function; it is a dilation from point $P$ followed by a translation by $\ub$. It could
perhaps be called a \emph{displacement function}.

Suppose $t \neq 0$. If we take $\ub = t(Q-P)$, then we can write the same displacement function in the \emph{root-slope form}
\begin{equation}
0 \mapsto  t(Q-O).
\end{equation}
The one exceptional case is when $t = 0$. Then there is no root (unless $\ub = \zerob)$, and the function may be identified with
the vector $\ub$.

While the above representation is appropriate for  calculation,
there is a  related vector field representation  that gives nice pictures.
Consider a function from $E$ to $E$ of the form
\begin{equation}
O \mapsto  O + t (P-O) + \ub = (1-t)O + t P + \ub.
\end{equation}
The graph of this function is a set of ordered pairs of points in $E$, and each ordered pair may be
thought of as a representation of a bound vector.
In particular, a weighted point $tQ$ is represented by the function
\begin{equation}
O \mapsto (1-t) O + t Q .
\end{equation}
If  $0<t< 1$, then the bound vector at $O$ goes part way from $O$ to $Q$. If $t=1$, then it goes all the way from $O$ to $Q$.
Before $t$ was a weight attached to $Q$. Here it is a measure of how much other points are attracted to $Q$.

\section{Representation of line bivectors as affine functions from $E$ to $\Lambda^2(V)$}

The space $\Lambda^2(V^\bullet)$ of line bivectors  also has an affine function model. It is realized as certain affine functions from $E$ to $\Lambda^2(V)$.
 Such a function is a \emph{moment function} of the form
\begin{equation}
O \mapsto   \Mb(O) = (P-O) \land \ub + \alpha,
\end{equation}
where $\ub$ is in $V$ and  $\alpha$ is in $\Lambda^2(V)$. This is a \emph{point-slope} representation of an affine function.

This is particularly easy to visualize in the
case $n=2$. Then the function is an affine function from the plane $E$ to a one-dimensional space, and every
affine function is of this form. In the cases $n \geq 3$ these functions are not so simple; they are affine
functions on $E$ with values in a space of dimension ${n \choose 2}$, and they are of a rather special form.

Here is an important remark about notation.
Even if we think of a line bivector as a moment function $O \mapsto (P-O) \land \ub + \alpha$, it is possible
to use the same notation as in the projective space representation. All one has to do is to regard the point $P$
as defining the function $O \mapsto P-O$. Then the moment function is $P \land \ub + \alpha$.

\begin{proposition}
A moment function $\Mb$ is
 determined by its values on any three non-collinear points in $E$.
 \end{proposition}

Consider three non-collinear points $O_1, O_2, O_3$. It is sufficient to show that if
$\Mb(O_1), \Mb_(O_2), \Mb(O_3)$ are all zero, then $\Mb = \zerob$. Write
$\Mb(O) = (P - O) \land \ub + \alpha$. Subtraction gives
$(O_2 - O_1) \land \ub = \zerob$ and $(O_3 - O_1) \land \ub = \zerob$. Since $O_2 - O_1$ and $O_3 - O_1$ are linearly
independent, the bivector lemma implies that $\ub = \zerob$.
It follows that $\Mb= \zerob$.

\begin{proposition} A moment function $\Mb$ is determined by its vector invariant $\ub$ together with its value $\Mb(O^*)$ on a single point.
\end{proposition}

This proposition says that given a point $O^*$ in $E$, there is a corresponding isomorphism
\begin{equation}
\Lambda^2( V^\bullet) \rightarrow V \oplus \Lambda^2(V)
\end{equation}
given by
\begin{equation}
P \land \ub + \alpha \mapsto (\ub, (P - O^*) \land \ub + \alpha).
\end{equation}
It is easy to see that given a pair $(\ub, \mu)$ one can recover a moment function
$O \mapsto  (O^* - O) \land \ub + \mu$.
This representation gives a rather concrete representation of line bivectors as vector, bivector pairs. Its drawback is that it is awkward to reason with an arbitrary choice of reference point.

The values $\Mb(O)$ of the moment function satisfy $\ub \land \Mb(O) = \ub \land \alpha = \kappa$, where $\kappa$ is the trivector invariant.
This says that these values lie in a hyperplane in the space of bivectors. This hyperplane intersects the origin only
when the trivector invariant is zero.
When the vector invariant is non-zero and the  trivector invariant is zero, then the affine function represents a sliding vector and may be written
\begin{equation}
O \mapsto  \Mb(O) = (Q-O) \land \ub .
\end{equation}
This is the \emph{root-slope representation} of the affine function.

\section{Scalar products of bivectors}

The  following sections exploit the fact that   the space $V$ and and the space $\Lambda^2(V)$ of bivectors over $V$ each have a scalar product (inner product, dot product).
With this extra structure
line bivectors can be given a more explicit form. This section presents the basic definitions.

For decomposable bivectors the scalar product is given by the determinant
\begin{equation}
(\vb \land \wb) \cdot (\xb \land \yb) = (\vb \cdot \xb )(\wb \cdot \yb) -
(\vb \cdot \yb) (\wb \cdot \xb) .
\end{equation}
This is zero if the two corresponding planes are orthogonal.
The \emph{magnitude} or \emph{area} of a decomposable bivector is
\begin{equation}
|\ub \land \wb| =   \area(\vb,\wb) = \sqrt{ (\vb \land \wb) \cdot (\vb \land \wb) }   .
\end{equation}

If $\eb_i$ for $i = 1, \ldots, n$ is an orthonormal basis for the $n$ dimensional space $V$, then the $\eb_i \land \eb_j$ for $1 \leq i < j \leq n$ is an orthonormal basis for the ${n \choose 2}$ dimensional space of bivectors. The general bivector is $\sum_{i<j} c_{ij} \eb_i \land \eb_j$. For $n=2$ the most general bivector is  $\alpha = c_{12} \eb_1 \land \eb_2$.  For $n=3$ the most general bivector is $\alpha =c_{23} \eb_2 \land \eb_3 + c_{13} \eb_1 \land \eb_3  + c_{12} \eb_1 \land \eb_2$.

There is an operation of \emph{interior product}. Let $\ub$ be a vector and let $\alpha$ be a bivector. Then the
interior product $\ub \interiorproduct \alpha$ is a vector. If $\alpha = \vb \land \wb$ is decomposable, then
the interior product is $\ub \interiorproduct (\vb \land \wb)= (\ub \cdot \vb) \wb - (\ub \cdot \wb) \vb$. This can easily
be remembered as having the formal structure of the product rule for differentiation, with the proviso that moving $\ub$ past $\vb$
introduces a minus sign. The condition that $\ub \interiorproduct (\vb \land \wb) = \zerob$ is equivalent to either $\ub$ being
orthogonal to both $\ub$ and $\wb$, or $\vb \land \wb = \zerob$.

\begin{lemma}  Fix a vector $\ub$. There are two maps
$\alpha \mapsto \ub \interiorproduct \alpha$ and $\vb \mapsto \ub \land \vb$; the nullspace of the first and the
range of the second are orthogonal complements.
Thus every bivector $\alpha$ may be written as an orthogonal sum $\alpha = \beta + \gamma$,
where $\ub \interiorproduct \beta = \zerob$, and $\gamma = \ub \land \zb$ for some vector $\zb$.
\end{lemma}

This can be proved as follows.
 Suppose $\ub \neq \zerob$.   Consider an orthonormal basis such that $\ub = t \eb_1$ with $t \neq 0$. Write $\alpha = \sum_{i<j} c_{ij} \eb_i \land \eb_j$.
 Suppose that $\ub \land \alpha = t \sum_{1<i<j} c_{ij} \eb_1 \land \eb_i \land \eb_j = \zerob$. If $\beta = \sum_{1 < i < j}  c_{ij} \eb_i \land \eb_j$,
 then $\ub \land \beta = \zerob$. Furthermore, $\gamma = \sum_{1 < j}  c_{1j} \eb_1 \land \eb_j = \ub \land \zb$, where
$\zb = (1/t) \sum_{1 < j}  c_{1j}  \eb_j$.

\begin{proposition} Consider a vector space $V$ with an inner product. Then the space $\Lambda^2(V)$ may be identified with the space of anti-symmetric
linear transformations on $V$. In other words, the elements of $\Lambda^2(V)$ belong to the Lie algebra of infinitesimal rotations.
\end{proposition}

This identification works as follows. The general non-zero decomposable bivector is a multiple of $\alpha =  \xb \land \yb$, where
$\xb$ and $\yb$ are orthogonal unit vectors.  Then $\ub \interiorproduct \alpha = (\ub \cdot \xb) \yb - (\ub \cdot \yb) \xb$. This should be
compared with the projection of $\ub$ on the plane, which is $P \ub = (\ub \cdot \xb) \xb + (\ub \cdot \yb) \yb$. Let $J$ be the quarter turn transformation of
the plane with $J \xb = \yb$ and $J \yb = - \xb$. Then $\ub \interiorproduct \alpha = J P \ub = PJP \ub$. Bivectors $\alpha$ of this form may be chosen
to form a basis for $\Lambda^2(V)$. The corresponding infinitesimal rotations $X = PJP$ form a basis for the Lie algebra. There is more on the Lie algebra point of view in the final section.

\section{The central axis and the bivector invariant}

The following result is a version of the Poinsot central axis theorem, after Louis Poinsot  (1777--1859).

\begin{theorem}
Every line bivector  $\Mb = P \land \ub + \alpha$ has a unique decomposition
\begin{equation}
\Mb =  Q \land \ub + \beta,
\end{equation}
with $\ub \interiorproduct \beta = \zerob$.
\end{theorem}

Proof: Write $\alpha = \beta + \gamma$ as
an orthogonal decomposition,
where $\ub \interiorproduct \beta = \zerob$ and  $\gamma = \wb \land \ub$.
So we can write the general line bivector as
\begin{equation}
P \land \ub + \alpha = P \land \ub + \wb \land \ub + \beta = (P + \wb) \land  \ub + \beta = Q \land \ub + \beta,
\end{equation}
with $Q = P +\wb$ and where $\ub \interiorproduct \beta = \zerob$.

The line $L$ determined by $Q$ and $\ub$ is called the \emph{central axis} of the line bivector. (When $\ub = \zerob$ the role of the central axis is
taken by the entire space $E$.) The line bivector then has two
parts, the sliding vector $Q \land \ub$ along the central axis, and the \emph{bivector invariant} $\beta$ orthogonal to $\ub$.
The  vector $\ub$ expresses a tendency to move along the central axis. The bivector $\beta$ expresses
an infinitesimal rotation about the central axis. The uniqueness of the decomposition demonstrates the
structure of a line bivector as the sum of two rather different objects. However, in computations it may be inconvenient to
make this decomposition explicit at each stage.

Consider the moment function representation of a line bivector.
The general non-constant line bivector function is of the form
\begin{equation}
O \mapsto \Mb(O) = (Q-O) \land \ub + \beta,
\end{equation}
where $\ub \interiorproduct \beta = \zerob$.
This could be called the \emph{minimizer-slope representation}  of the affine function. The point $Q$ is
on the central axis. At such points the magnitude of $\Mb(O)$ assumes its minimum value

\section{Force and torque}

A common application of the theory is to force and torque. (This is usually in the case of dimension $n=3$.) Force is represented by a vector $\ub$.
Torque is given by a bivector $\alpha$ (often represented by pseudovector).
A line bivector is called a  \emph{wrench}. It is given by a moment function $O \mapsto \Mb(O)  = (P-O) \land \ub + \alpha$. This is a function defined on $E$ with values that are bivectors. It may be denoted in abbreviated form by
$\Mb = P \times \ub + \alpha$.

In three dimensions a  bivector $\alpha$ is always decomposable.
Such a bivector often arises as a \emph{force couple} given by a pair of opposite forces at different points in this plane.
 According to the Poinsot theorem, the wrench has a standard form $\Mb = Q \land \ub + \beta$.
 It has a torque part  given by a force $\ub$ acting at a point $Q$ in $E$ on
 the central axis. There is also a pure torque part $\beta$, a force couple in the plane orthogonal to $\ub$.
 This force couple is not associated with a location in $E$.

 In a typical rigid body static problem there are forces $\ub_1, \ldots, \ub_k$ applied at points $P_1, \ldots, P_k$. The sum of the
 corresponding sliding vectors is $P_1 \land \ub_1 + \cdots + P_k \land \ub_k$. Such a sum does not have to be a sliding vector.
 However the sum is always a line bivector, defined by a moment function. Furthermore, the total force associated with this
 line bivector is $\ub = \ub_1 + \cdots + \ub_k$. So it can be written
 \begin{equation}
 P_1 \land \ub_1 + \cdots + P_k \land \ub_k = P \land \ub + \beta,
 \end{equation}
 where   $P$ is on the central axis, and where the force couple $\beta$ is a torque in the plane orthogonal to the force $\ub$.

In the special case when all the points are in the same plane and all the forces act in this plane, it is impossible to have
the force couple $\beta$ orthogonal to the force $\ub$, unless either $\beta = \zerob$ or $\ub = \zerob$. For a planar
problem the line bivector is either a sliding vector or it is a force couple.

In general the sum of sliding vectors is a line bivector, and the corresponding moment function can depend
on the reference point. However the condition for equilibrium is that the sum of all the sliding vectors is the zero line bivector; for
 this special case the moment function is independent of the reference point.
Thus an  equilibrium calculation has no need of a particular reference point. The natural framework for
such a calculation is Grassmann algebra, in which the exterior product of a point and a vector is a sliding vector.

\section{Duality}

This last section presents a somewhat different point of view on line bivectors. This framework is particularly appropriate for mechanics,
since it takes into account the fact that displacements, forces, and torques  have different units. It is helpful to distinguish a vector
space from its dual vector space, even though they have the same dimension. We shall see that line bivectors then have natural dual objects.

Again $E$ is $n$-dimensional Euclidean space, an affine space.
The space $V$ is the $n$-dimensional real vector space of free vectors, and $V^*$ is its $n$-dimensional dual vector space. This dual space consists
of real linear functions on $V$. Let $L$ be the Lie algebra of the rotation group of $V$. This is a vector space of dimension ${n \choose 2}$
consisting of anti-symmetric linear transformations from $V$ to $V$. Let $L^*$ be the dual vector space of $L$. Each of these vector spaces has a physical
interpretation. An element of $V$, $V^*$, $L$, or $L^*$ is a displacement, force, angular velocity, or torque.

Suppose that $\zb$ is in $V$ and $\ub$ is in $V^*$. The corresponding \emph{moment} (or torque) $\zb \sqcap \ub$ is an element of $L^*$. For $\omega$ in $L$, this
is given by
\begin{equation}
(\zb \sqcap \ub) \omega = \ub \omega \zb .
\end{equation}
Here $\omega \zb$ is in $V$, so the right hand side is a scalar. The map from phase space $(\zb, \ub)$ to $\zb \sqcap \ub$ in $L^*$ is often called the
\emph{moment map}. (The notation $\zb \sqcap \ub$ is adopted here to make explicit the analogy with exterior product.)
See  Abraham and Marsden (1978) for much more on this group theoretic approach to mechanics.

Consider a point $P$ in $E$ and an element $\ub$ in $V^*$ and a corresponding $\Mb(P)$ in $L^*$. The corresponding
moment function $\Mb$ is a function from $E$ to $L^*$ given by
\begin{equation}
\Mb(O) = (P - O) \sqcap \ub + \Mb(P).
\end{equation}
This definition is a repackaging of the previous moment function definition of line bivector. In the application where $\ub$ is force,
it is natural to call this a \emph{wrench}.

The  dual to a wrench is a quite different object. Let $V'$ be another copy of $V$, but interpret an element of $V'$ as a velocity instead of a
displacement. Consider a point $Q$ in $E$ and an element $\omega$ in $L$ and a vector $\vb(Q)$ in $V'$. Define a function $\vb$ from $E$ to $V'$ by
\begin{equation}
\vb(O) = \omega (O-Q) + \vb(Q).
\end{equation}
Such a function may be called a \emph{twist}.
 (When $n=3$ it is easy to confuse a twist with a wrench, but for general $n$ not even
the dimensions match up.)

There is a duality relation between a wrench and a twist. The \emph{bilinear scalar invariant} is $\ub \vb(O) + \Mb(O) \omega$.
Using the definition of moment, this works out to be
\begin{equation}
\ub \vb(O) + \Mb(O) \omega = \ub \vb(Q) +  \ub \omega (P-Q)   +          \Mb(P) \omega .
\end{equation}
This is independent of the reference point $O$. In particular, the bilinear scalar invariant may be
written $\ub \vb(P) + \Mb(P) \omega = \ub \vb(Q) + \Mb(Q) \omega$.

One important physical application is to forces acting on a rigid body in motion. The wrench $\Mb(O) = \sum_{i=1}^k    (P_i - O) \sqcap \ub_i $ is a sum of sliding vectors. Each term represents a force $\ub_i$ acting at point $P_i$. The velocity at point $P_i$ is
$ \vb(P_i) = \omega (P_i - O) + \vb(O)$.  The \emph{power} is $\sum_{i=1}^k  \ub_i \vb(P_i)$. There is a corresponding power theorem.

\begin{theorem}
Power is the bilinear scalar invariant associated with wrench and twist.
\end{theorem}

The computation of the power begins
\begin{equation}
\sum_{i=1}^k  \ub_i \vb(P_i) = \sum_{i=1}^k  \ub_i \omega (P_i - O)  + \sum_{i=1}^k  \ub_i \vb(O).
\end{equation}
Using the definition of  moment  this is
\begin{equation}
\sum_{i=1}^k  ((P_i - 0) \sqcap \ub_i)  \omega + \sum_{i=1}^k  \ub_i \vb(O) = \Mb(O) \omega + \ub \vb(O).
\end{equation}

The scalar product on $V$ allows an  element  of the Lie algebra to be represented by a bivector. In
dimension $n=3$ the bilinear scalar invariant of this section
is closely related to the bilinear trivector invariant considered before. However, this is special to
three dimensions, and it helps to look at at general $n$ to fully appreciate the relation  between wrench, twist, and power.

\end{document}